\input harvmac

%%%%%%%%%%%%%%%%%%
\Title{\vbox{\baselineskip12pt
\hbox{CERN-TH/96-58}
\hbox{McGill/96-06}
\hbox{hep-th/9602153}
}}
{\vbox{\centerline{Strings from Membranes and Fivebranes}}}

\centerline{Ramzi R.~Khuri\footnote{$^*$}{Based in part on talk given
at the
Workshop on Recent Developments in Theoretical Physics:
``STU-Dualities and Non-Perturbative Phenomena in Superstrings and
 Supergravity'', held at CERN November 27th-December 1st 95, and to
appear
in {\bf Progress in Physics}. Supported by a World Laboratory
Fellowship.}}
\bigskip\centerline{{\it Theory Division, CERN,
 CH-1211, Geneva 23, Switzerland}}
\medskip\centerline{and}
\medskip\centerline{{\it Physics Department, McGill
University, Montreal, PQ, H3A 2T8 Canada}}
\vskip .3in

Under the six-dimensional heterotic/type $IIA$ duality map,
a solitonic membrane solution of heterotic string theory transforms
into
a singular solution of type $IIA$ theory, and should therefore be
interpreted
as a fundamental membrane in the latter theory. This finding pointed
to a gap
in the formulation of string theory that was subsequently filled by
the
discovery of the role of $D$-branes as the carriers of Ramond-Ramond
charge in type $II$ string theory. The roles of compactified
eleven-dimensional membranes and fivebranes in five-dimensional
string theory
are also discussed.

\vskip .3in
\Date{\vbox{\baselineskip12pt
\hbox{CERN-TH/96-58}
\hbox{McGill/96-06}
\hbox{February 1996}}}

\def\sqr#1#2{{\vbox{\hrule height.#2pt\hbox{\vrule width
.#2pt height#1pt \kern#1pt\vrule width.#2pt}\hrule height.#2pt}}}

\def\a{\alpha}

\def\ie{{\it i.e.,}\ }

\def\[{\bf (*}
\def\]{*) \rm\ }

\lref\vafawit{C.~Vafa and E.~Witten, Nucl. Phys. {\bf B447} (1995)
261.}

\lref\stst{M.~J.~Duff and R.~R.~Khuri, Nucl. Phys. {\bf B411}
(1994) 473.}

\lref\duflufb{M.~J.~Duff and J.~X.~Lu, Nucl. Phys. {\bf B354} (1991)
141;
C.~G.~Callan, J.~A.~Harvey and A.~Strominger, Nucl. Phys. {\bf B359}
(1991) 611.}

\lref\duffsw{M.~J.~Duff, Nucl. Phys. {\bf B442} (1995) 47.}

\lref\duffm{M.~J.~Duff and R.~Minasian, Nucl. Phys. {\bf B436} (1995)
507.}

\lref\dufflm{M.~J.~Duff, J.~T.~Liu and R.~Minasian, Nucl. Phys.
{\bf B452} (1995) 261.}

\lref\witfb{E.~Witten, hep-th/9512062.}

\lref\duffmw{M.~J.~Duff, R.~Minasian and E.~Witten, hep-th/9601036.}

\lref\hult{C.~M.~Hull and P.~K.~Townsend, Nucl. Phys. {\bf B438}
(1995) 109.}

\lref\wit{E.~Witten, Nucl. Phys. {\bf B443} (1995) 85.}

\lref\schwarz{J.~H.~Schwarz, hep-th/9510086; hep-th/9601077.}

\lref\horw{P.~Ho\v rava and E.~Witten, Nucl. Phys. {\bf B460} (1996)
506.}

\lref\antft{I.~Antoniadis, S.~Ferrara and T.~R.~Taylor, Nucl. Phys.
{\bf B460} (1996) 489.}

\lref\becker{K.~Becker, M.~Becker and A.~Strominger, Nucl. Phys.
{\bf B456} (1995) 130.}

\lref\prep{M.~J.~Duff, R.~R.~Khuri and J.~X.~Lu,
Phys. Rep. {\bf 259} (1995) 213.}

\lref\sing{M.~J.~Duff, R.~R.~Khuri and J.~X.~Lu,
Nucl. Phys. {\bf B377} (1992) 281.}

\lref\hmono{R.~R.~Khuri, Phys. Lett. {\bf B259} (1991) 261;
Nucl. Phys. {\bf B387} (1992) 315.}

\lref\dufkmr{M.~J.~Duff, R.~R.~Khuri,
R.~Minasian and J. Rahmfeld, Nucl. Phys. {\bf B418} (1994) 195.}

\lref\hosono{S.~Hosono, A.~Klemm, S.~Theisen and S.~T.~Yau,
Commun. Math. Phys. {\bf 167} (1995) 301;
S.~Kachru and C.~Vafa, Nucl. Phys. {\bf B450} (1995) 69.}

\lref\town{P.K.~Townsend, Phys. Lett. {\bf B350} (1995) 184;
Phys. Lett. {\bf B354} (1995) 247; C.M.~Hull and P.K.~Townsend,
Nucl. Phys. {\bf B451} (1995) 525.}

\lref\towns{P.~K.~Townsend, hep-th/9507048.}

\lref\gibbons{G.~W.~Gibbons, Nucl. Phys. {\bf B207} (1982) 337;
D.~Garfinkle, G.~T.~Horowitz and A.~Strominger,
Phys. Rev. {\bf D43} (1991) 3140; erratum {\bf D45} (1992) 3888.}

\lref\solitons{A.~Sen, Nucl. Phys. {\bf B450} (1995) 103;
J.~A.~Harvey and A.~Strominger, Nucl. Phys. {\bf B449} (1995) 535.}

\lref\strom{A.~Strominger, Nucl. Phys. {\bf B343} (1990) 167.}

\lref\klopv{R.~Kallosh, A.~Linde, T.~Ortin, A.~Peet and
A.~Van Proeyen, Phys. Rev. {\bf D46} (1992) 5278.}

\lref\mont{C.~Montonen and D.~Olive, Phys. Lett. {\bf 72B}
(1977) 117.}

\lref\poly{A.~M.~Polyakov, Phys. Lett. {\bf 103B} (1981) 207,
211.}

\lref\tension{A.~Dabholkar and J.~A.~Harvey, Phys. Rev. Lett.
{\bf 63} (1989) 478; A.~Dabholkar, G.~Gibbons, J.~A.~Harvey and
F.~Ruiz
Ruiz, Nucl. Phys. {\bf B340} (1990) 33.}

\lref\jkkm{C.~V.~Johnson, N.~Kaloper, R.~R.~Khuri and R.~C.~Myers,
Phys. Lett. {\bf B368} (1996) 71.}

\lref\polchinski{J.~Polchinski, Phys. Rev. Lett. {\bf 75} (1995)
4724.}

\lref\fkm{S.~Ferrara, R.~R.~Khuri and R.~Minasian, hep-th/9602102.}

%%%%%%%%%%%%%%%%%%%%%%%%%%%%%%%%%%%%%%%%%%%%%%%%%%%%%%%%%%%%%%%%

Low energy string theories contain a rich
array of solutions corresponding to extended
objects, the so-called $p$-branes
(see \prep\ and references therein).
It is now apparent that
these objects play an important role in the non-perturbative
physics of string theories \becker.
To this end, a fundamental formulation
of string theory that goes beyond the first quantized framework of
the
Polyakov path integral \poly\ is required. Progress into
understanding the strong coupling dynamics of certain supersymmetric
string
theories has been made in \wit, and which may provide new insights
into the
correct fundamental framework which must underly string theory.

This talk is divided into two parts. In the first, we follow \jkkm\
and present evidence that type $IIA$ superstrings are only one
component of a
larger theory which also contains fundamental membranes, a finding
that
foreshadowed the subsequent discovery of the role of Dirichlet-branes
($D$-branes) as the carriers of Ramond-Ramond charge in type $II$
string theory \polchinski.
In the second part, we summarize the recent results of \fkm, in which
a five-dimensional string is obtained from
eleven-dimensional membrane and fivebrane solitons wrapped around
two-cycles and four-cycles of a Calabi-Yau threefold.

According to string/string duality \refs{\stst,\hult,\wit}, the
strong
coupling physics of certain superstring
theories may be reformulated as the weak coupling
physics of ``dual'' string theories.  One interesting example
is the duality in six dimensions between heterotic strings
compactified on $T^4$ and type $IIA$ superstrings compactified on
$K3$.
In fact, the duality relies on the much stronger conjecture that
these two strings are completely equivalent
\refs{\hult,\wit,\vafawit}.
One further result which supports the equivalence
is that the heterotic string can be identified
as a soliton within the type $IIA$ string theory, and conversely,
the type $IIA$ string can be identified
as a soliton in heterotic string theory \solitons.
Thus under the duality
transformation, the roles of the fundamental and solitonic strings
are interchanged. This interchange is a stringy version of the
role reversal between magnetic monopoles and electric charges
arising in the strong/weak coupling duality of
gauge field theories \mont.

For a given $p$-brane solution of a string theory, the question
arises as to
how this solution behaves under the strong/weak coupling
duality transformations discussed above.
There are three distinct possibilities: (i) the $p$-brane could be
a singular field configuration in both of the dual string theories,
which would justify discarding it
as unphysical, (ii) the $p$-brane could be nonsingular in both
theories,
in which case it would be treated as a soliton in both contexts,
and finally (iii) the $p$-brane could be nonsingular in one
theory but singular in the dual theory. In the latter case, since
it appears as a soliton in one theory, one would not be able to
omit it from the spectrum. However the fact that the $p$-brane
solution is singular in the dual theory suggests that it represents
the external fields around a fundamental
source\foot{In case (i), one could
also consider the possibility that the $p$-brane
is fundamental in both of the theories.} -- \ie the dual
theory should contain fundamental $p$-branes!

The singularity structure of a solution is determined by
examining the $p$-brane with a certain test-probe,
\ie determining the behavior of a small test object as it approaches
the core of the $p$-brane. The choice of the test-probe would depend
on which fundamental theory underlies the original brane solution
\refs{\sing,\prep}. This amounts to measuring possible curvature
singularities with the metric which couples to the world-volume
of the fundamental objects in the theory, \ie the
metric which appears in the sigma-model describing these
fundamental objects. For example, in
heterotic string theory, the natural test-probe to examine
a $p$-brane solution would be a fundamental heterotic string.
Applied to the case of the six-dimensional string/string duality,
this
means that the heterotic string appears
singular in the heterotic string sigma-model
metric, but is nonsingular in the type $IIA$ superstring metric
\solitons.

Consider heterotic string theory compactified on a torus down to six
dimensions. For a generic point in the moduli space, the low
energy effective theory is $N=2$ supergravity coupled to
twenty abelian vector multiplets.
Thus the bosonic fields include the metric, the dilaton, the
antisymmetric Kalb-Ramond field, $24$ abelian gauge fields, and
$80$ scalar moduli fields.
In six dimensions the three-form field strength of the
Kalb-Ramond two-form couples naturally as the ``electric'' or
``magnetic'' field around a one-brane, or string.
In fact these correspond to the two string solutions discussed above,
\ie the fundamental heterotic string with the electric
Kalb-Ramond charge, and its dual solitonic string, with
the magnetic three-form charge. Point-like or zero-brane solutions
also appear, with conventional electric charges from the $U(1)$
two-form field strengths.
In particular, singular point-like objects arise
as the extremal limits of electrically charged black holes.
In this case, the dual
objects are two-branes, or membranes, with magnetic $U(1)$ charge.
To complete the list, one could also consider three-branes
which carry a ``magnetic'' charge from the periodic moduli scalars,
and ``minus-one''-branes or instantons carrying scalar electric
charge.
We restrict our attention, though, to a class of solitonic
membranes.

It is consistent to truncate the low energy action to the form
\eqn\hetact{S_{het}=\int d^6\!x\, \sqrt{-G}\, e^{-2\Phi}
\left(R + 4(\partial\Phi)^2 - {1\over 4}F^2\right),}
where $F=dA$ is the field strength for one of
the $U(1)$ gauge fields, $\Phi$ is the six-dimensional dilaton
and $G_{\mu\nu}$ is the heterotic string sigma-model metric.
For this action, one finds the following solution
which represents a magnetically charged membrane
\eqn\hetmem{\eqalign{ds^2&=-dt^2+dx_1^2+dx_2^2
+\left(1+{Q\over y}\right)^2\left(dy^2+y^2d\Omega_2^2\right),\cr
e^{2\Phi}&=1+{Q\over y}, \cr
F_{\theta\varphi}&=\sqrt{2}Q\sin\theta.\cr}}
Here $(y,\theta,\varphi)$ are polar coordinates on the
$(x_3,x_4,x_5)$
subspace,
and $d\Omega_2^2$ is the line element on the unit
two-sphere\foot{This
solution is simply the magnetically-charged
extreme dilaton black hole from four dimensions \gibbons,
raised to six dimensions by adding the flat $x_1,x_2$ directions,
which are tangent to the membrane.}.

While the metric in \hetmem\
may appear singular at the core of the membrane,
this is a coordinate artifact. In fact, the
solution develops an infinitely long throat with a constant
radius
as $y\to 0$, as is most easily recognized with the coordinate
transformation $\rho/Q=\log(y/Q)$. Then the fields near the core
become
\eqn\throatsol{\eqalign{ds^2&\simeq
-dt^2+dx_1^2+dx_2^2+d\rho^2+Q^2\,d\Omega^2_2,\cr
\Phi&\simeq-\rho/2Q, \cr
F_{\theta\varphi}&=\sqrt{2}Q\sin\theta.\cr}}
The above description of the throat geometry is made using
the heterotic string sigma-model metric, and hence this
membrane is completely nonsingular for the heterotic string
test-probes.

We are interested in considering this solution in the
strong coupling regime in which the dual type $IIA$ string theory
is weakly coupled. Thus we seek a supersymmetric
membrane saturating a BPS bound, for which the mass-charge
relation is preserved against higher-order corrections
in the strong coupling regime \tension.
Therefore, we choose as our gauge field
one of the four contained in the supergravity multiplet. This
provides
a supersymmetric embedding of \hetmem\ in the full six-dimensional
$N=2$ theory in which half of the spacetime supersymmetries are
preserved \klopv.
In ten-dimensional heterotic string theory,
this choice of gauge fields corresponds to setting
$G_{i\mu}=B_{i\mu}=A_\mu$, where $G$ and $B$ denote the
ten-dimensional
metric and Kalb-Ramond field, respectively, with $\mu=0,1,2,3,4,5$,
a spacetime index
and $i=6,7,8$ {\it or} 9 corresponding to {\it one} of the directions
compactified on the four-torus. The ten-dimensional throat solution
is then
a constant radius three-sphere supported by the parallelizing
torsion of the Kalb-Ramond field, a linear dilaton background
in the $\rho$ direction, and five flat spatial directions and
a trivial time direction. This corresponds precisely to the
throat limit of the ten-dimensional neutral fivebrane solution
\duflufb,
and so reveals that our membrane is in fact a fivebrane ``warped''
around the toroidally compactified directions\foot{The
solution is a ``warped" as opposed to ``wrapped" fivebrane
\refs{\strom,\stst}. The latter
dimensionally reduces to an $a= \sqrt{3}$
black hole/$H$-monopole \hmono\ in $D=4$
as opposed to the $a=1$ solution we started
with in this paper. In our warped solution one of the compact
directions is tied up in the three-sphere
surrounding the fivebrane in a topologically nontrivial way.}.
The throat solution is essentially unchanged, and one
is guaranteed that no singularities develop at the membrane core
\jkkm.
Thus despite the appearance of $\a'$ corrections, we are assured
that the membrane is a stable soliton of the heterotic string
We also expect that the background Killing spinors are perturbatively
corrected so that spacetime supersymmetry also survives the $\a'$
corrections \jkkm.

We now transform the membrane soliton to the type $IIA$
string theory via the duality mapping \wit
\eqn\duality{\Phi'=-\Phi, \qquad\qquad
G_{\mu\nu}'=e^{-2\Phi} G_{\mu\nu},\qquad\qquad
A_\mu'=A_\mu.}
Here the (un)primed fields are those arising in the type $IIA$
(heterotic) string theory. In particular, $G_{\mu\nu}'$
is the metric which couples to the type $IIA$ string sigma-model.
The type $IIA$ action is then given by
\eqn\iiact{S_{IIA}=\int d^6\!x\, \sqrt{-G'} \left[e^{-2\Phi'}
\left(R' + 4(\partial\Phi')^2\right) -
{1\over 4}F'^2\right],}
and the solution becomes
\eqn\iimem{\eqalign{ds'^2&=\left(1+{Q\over y}\right)^{-1}
\left(-dt^2+dx_1^2+dx_2^2\right)
+\left(1+{Q\over y}\right)\left(dy^2+y^2d\Omega_2^2\right),\cr
e^{2\Phi'}&=\left(1+{Q\over y}\right)^{-1}, \cr
F_{\theta\varphi}'&=\sqrt{2}Q\sin\theta.\cr}}
In this frame the leading order solution becomes singular,
requiring a source to support it at the core. First, the core,
{\it i.e.,}\ $y=0$, is  a finite proper distance away, and the
curvature
diverges there, {\it e.g.,}\ the Ricci scalar goes as $R\sim 1/(Qy)$.
Thus from the point of view of type $IIA$
string test probes, the membrane appears singular.
Essentially with \duality, we have made a singular conformal
transformation of the original metric which implicitly
adds an extra ``point-at-infinity'' closing off the end of the
throat.
To consistently solve the new equations of motion for \iiact, we must
now include a source at this end-point, \ie $y=0$.
Hence in the type $IIA$ theory, the membrane must be interpreted
as fundamental.

So the nonsingular supersymmetric
solution in the heterotic string theory is singular in type
$IIA$ theory. Because of the nonsingular nature
of the solution, it appears that these field configurations
must be included in defining the heterotic string
theory. From this and related results, three possible alternatives to
describe the complete type $IIA$ theory were suggested in \jkkm:

{\it All-branes: an egalitarian theory of branes:}\/--
In this, the simplest alternative, the full type $IIA$ theory
is a theory which contains (at least) two distinct fundamental
objects, strings and membranes. First quantization would be
separately applied for each brane with its distinct world-volume
action. A second step would be to incorporate interactions between
the different branes in this first quantized framework.
Presumably in this theory, the membranes would not contribute
to the massless spectrum at a generic point in the
(known) vacuum moduli space,
since the latter spectrum is fully accounted for by type $IIA$
strings.
In this case, the membranes would play no role
in the low energy physics, but would be important for a consistent
definition of
the theory at the level of massive modes and through nonperturbative
effects \becker. Such an egalitarian description the type $IIA$
theory
was advocated in
\towns, where in fact on the basis of $U$-duality
the democracy was extended to all $p$-branes appearing in the
low energy theory.

{\it Big-branes: a theory of only higher branes:}\/--
In this second scenario, the true type $IIA$ theory would actually
be a theory of only membranes (or some higher $p$-branes).
The fundamental strings would then be ``string-like''
excitations of the membrane. In order for this alternative to be
consistent, the membranes must also be able to act as sources
for the Kalb-Ramond fields that are associated with the
fundamental type $IIA$ string.  This requirement could be confirmed
by
examining the zero-mode structure of these solutions.
Further, a much more stringent constraint is that
consistently quantizing the
fundamental membranes must reproduce precisely
the same massless spectrum as the type $IIA$ string in this $K3$
context. A higher brane
description of the type $IIA$ theory
was advocated in
\town\ with the suggestion that the correct
fundamental theory was an eleven-dimensional supermembrane theory.

{\it Something else:}\/--
On this alternative, little was said in \jkkm.
However we note that past efforts at quantizing higher $p$-branes
have
met with no success. Further even if a free first-quantized theory
was constructed, the introduction of interactions for higher
$p$-branes would remain a significant challenge. These technical
obstructions lend favor to the opinion that only one-branes, or
strings,
should be treated as fundamental. The present analysis,
which indicates that the type $IIA$ theory must incorporate
fundamental membranes, may then be an indication that the
correct fundamental description of the theory is simply not one based
on the first quantization of extended objects.

Interestingly, soon after \jkkm\ appeared, Polchinski \polchinski\
came up
with the proposal that
Dirichlet-branes ($D$-branes), extended objects defined by
mixed Dirichlet-Neumann boundary conditions, are the carriers of
electric
and magnetic Ramond-Ramond charge. Once $D$-branes are added as
Ramond-Ramond sources to type $II$ string theory,
the questions raised above are effectively answered. In particular,
as a carrier of Ramond-Ramond charge, the membrane constructed above
has the interpretation of a $D$-brane, and its mass per unit area
has the correct dependence on the string coupling constant \jkkm. As
$D$-branes
will appear throughout these proceedings, we will not discuss them
here
in any detail. Suffice to say that their discovery fits in nicely
with the simple physical picture described above in
showing that, if string/string duality is to be taken seriously, the
present
formulation of string theory as a theory of only strings is
insufficient, and
that fundamental membranes are required to couple in some manner in
order
to complete the picture.

Recent activity has also focused on the conjecture of the existence
of an
underlying eleven-dimensional theory (the so-called $M$-theory
\refs{\wit,\dufflm\schwarz\horw\witfb{--}\duffmw}),
whose low-energy limit is eleven-dimensional supergravity. $M$-theory
also
clearly fits in with the above discussion, and its
eventual construction should lead
to the establishment of the various string/string dualities
\refs{\stst,\duffm,\hult,\wit,\duffsw,\duffmw}.
In this framework, the five seemingly distinct string theories
arise as weak coupling limits of the various compactifications of the
eleven-dimensional $M$-theory, in which the membrane and fivebrane
that
naturally arise are either wrapped around or reduced on the
compactified
directions. In the rest of this talk, we summarize the recent results
in
\fkm, in which evidence is presented
for a five-dimensional duality between $M$-theory
compactified on a Calabi-Yau threefold and heterotic string theory
compactified
on $K3\times S^1$.

In \antft, the conjecture was made that the effective theory of
heterotic string theory compactified on $K3\times S^1$ is dual to
eleven-dimensional supergravity compactified on a Calabi-Yau
threefold.
This theory is also equivalent to type $IIA$ string theory
compactified
on the same Calabi-Yau threefold, in an appropriate large volume
limit.
Point-like (electric) states are obtained in $D=5$ by wrapping the
membrane from $M$-theory around two-cycles in
the Calabi-Yau space. Denote two-cycles and four-cycles respectively
by $C^{2\Lambda}$ and $C_{4\Lambda}$, where
$\Lambda=1,...,h_{(1,1)}$.
The charges of these states are obtained from the charge of the
membrane by
\eqn\emem{e_\Lambda=\int_{C_{4\Lambda}\times S^3} G_7,}
where $G_7=\delta {\cal L}/\delta F_4$, where $F_4=dA_3$ is the
field strength of the three-form antisymmetric tensor field.
String-like (magnetic) states in $D=5$ arise by wrapping the
fivebrane
around four-cycles in the Calabi-Yau space. The charges of these
states
are then obtained from the charge of the fivebrane by
\eqn\mfiv{m^\Lambda=\int_{C^{2\Lambda}\times S^2} F_4.}
Since the membrane and fivebrane are electric/magnetic duals in
eleven
dimensions, the above point-like and string-like states are dual to
each
other in the electric/magnetic sense and correspond to point-like and
string-like soliton solutions \hult. Following the singularity
structure
criteria used to discuss the membrane above, one can show that
in $D=5$, each object is self-singular and mutually non-singular with
its dual.

In a recent paper \duffmw, heterotic string/string duality was
examined from the point of view of $M$-theory, where it was argued
that the $E_8\times E_8$ heterotic string compactified on $K3$ with
equal instanton
 numbers in the two $E_8$'s is
self-dual, a result that can be seen by looking in two different ways
at
eleven-dimensional $M$-theory compactified on $K3\times S^1/Z_2$.
One weakly coupled heterotic string is obtained by wrapping the
$D=11$
membrane around
$S^1/Z_2$, while the dual heterotic string, also weakly coupled, is
obtained by reducing the
$D=11$ fivebrane on $S^1/Z_2$ and then wrapping around $K3$. Each of
these
two strings is strongly coupled from the point of view of the dual
one.
If we further compactify by reducing the first six-dimensional
heterotic string on $S^1$ and wrapping the dual six-dimensional
heterotic
string on $S^1$, we obtain on the one hand a string in five
dimensions
and on the other a dual, point-like object in five dimensions.
We claim \fkm\ that, starting with a $K3$ vacuum in which the gauge
symmetry is
completely Higgsed,
this $D=5$ string can be identified with the $M$-theory
fivebrane wrapped around a Calabi-Yau four-cycle, while the $D=5$
point-like
object can be identified with the $M$-theory membrane wrapped
around a Calabi-Yau two-cycle for the specific Calabi-Yau manifold
$X_{24}(1,1,2,8,12)$ with $h_{(1,1)}=3$ and $h_{(2,1)}=243$ \hosono.
In five dimensions, this model
contains $n_V=h_{(1,1)}-1=2$ vector multiplets (not counting the
graviphoton) and $n_H=h_{(2,1)}+1=244$
hypermultiplets.\foot{Here we do not consider the hypermultiplet
sector of $M$-theory where the low-energy effective action in $D=5$
does receive
membrane and fivebrane instanton corrections \becker.}

It is straightforward to match the perturbative and non-perturbative
BPS states
arising from the ten-dimensional compactification with the states
displayed
in the previous section and arising
from the eleven-dimensional compactification.
{}From the ten-dimensional point of view,
the heterotic string compactified on $K3\times S^1$ has the
perturbative
fundamental string state with charge
\eqn\fundst{m_0=\int_{K3\times S^1\times S^2} H_7,}
where $H_7=e^{-\phi} *H_3$, $H_3$ is the field strength of the
two-form
antisymmetric tensor field and $\phi$ is the ten-dimensional dilaton.
This state has mass per unit length $M_0=m_0g_5^2$.
Here the string is not
wrapped around the $S^1$. The corresponding classical
solution is given by the fundamental string of \tension.
This mass formula,
which follows from central charge/supergravity considerations
\antft, can also be
obtained by computing the ADM mass of the fundamental string
solution. This state is associated with the $b_{\mu\nu}$ field and is
dual to a vector in $D=5$.
The string theory also possesses
a perturbative electrically charged
point-like $H$-monopole state
(dual to the magnetically charged $H$-monopole state
of \hmono) with charge
\eqn\phmono{e_1=\int_{K3\times S^3} H_7}
and with mass $M_1=e_1Rg_5$, where $R$ is the radius of the $S^1$ and
$g_5$
is the five-dimensional string coupling constant.
In this case, the string is wrapped around the $S^1$.
Again one obtains the
same mass from either the central charge or the ADM mass of the
solitonic
solution. This state is associated
with the $b_{\mu 6}$ field. The $T$-dual electrically charged
point-like
Kaluza-Klein state with charge $e_2$ and associated with the
$g_{\mu 6}$ field has mass $M_2=e_2g_5/R$. In this case, the
corresponding
electrically charged solution is given by the extremal Kaluza-Klein
black
hole solution of heterotic string theory \dufkmr.
The fundamental
string state can be identified with one of the three states arising
from
the $M$-theory fivebrane, while the $H$-monopole and Kaluza-Klein
states
can be identified with two of the three states arising from the
$M$-theory
membrane.

The dual case is similar:
the heterotic fivebrane wrapped around $K3\times S^1$ has the
non-perturbative
(from the string point of view) point-like state with charge
\eqn\fundpt{e_0=\int_{S^3} H_3}
and mass $M'_0=e_0/g_5^2$ \refs{\wit,\antft}. Here the classical
solution is
simply the heterotic fivebrane of \duflufb\ wrapped around $K3\times
S^1$,
and which is dual to the fundamental heterotic string.
One also gets from the heterotic
fivebrane a non-perturbative magnetically charged
string-like $H$-monopole state with charge
\eqn\shmono{m_1=\int_{S^1\times S^2} H_3}
and mass per unit length $M'_1=m_1R/g_5$, where in this case the
fivebrane is
wrapped around the $K3$ but reduced on the $S^1$.
The solution in this case is the usual magnetically
charged $H$-monopole, which in $D=5$ is a string \hmono.
The $T$-dual magnetically charged string-like
Kaluza-Klein state with charge $m_2$ has mass per unit length
$M'_2=m_2/g_5R$.
The point-like state can be identified with one of the three states
shown
in the previous section arising from the $M$-theory membrane, while
the
string-like $H$-monopole and Kaluza-Klein states
can be identified with two of the three states shown in the previous
section arising from the $M$-theory fivebrane.

Note that each of the
three pairs of electric/magnetic dual states obey Dirac quantization
conditions. Note also that neither the membrane nor the fivebrane
from
$M$-theory is in itself sufficient to reproduce the perturbative
spectrum
of either the five-dimensional string or the dual five-dimensional
point-like
object. This becomes clear when one realizes that, from the
$M$-theory side,
the membrane wrapped around a two-cycle yields only point-like
states, while
the fivebrane wrapped around a four-cycle yields only string-like
states. On
the other hand, from the heterotic compactification,
both the string and point-like theories in $D=5$
contain both string and point-like objects in their perturbative
spectra.
In particular, it follows that the $D=5$ spectrum of Calabi-Yau
string
solitons yields the fundamental string states on the heterotic side
as well
as the non-perturbative heterotic string states obtained by wrapping
the
heterotic fivebrane on $K3$.

One-loop calculations providing further evidence for this duality
were
shown in \fkm. It was also found that, from anomaly considerations, a
five-dimensional string action arises which is chiral
on the worldsheet. This is especially interesting, since it implies
that $M$-theory
calculations may be carried out in the more familiar setting of
string theory. Further reduction to $D=4$
yields the standard dual $N=2$ supersymmetric theories,
but one may hope to obtain dual $N=1$ chiral theories
following \duffmw\ by considering two different limits of $M$-theory
compactified on $CY \times S^1/Z2$.

{\bf Acknowledgements}

This contribution is based mainly on two collaborations, one with
Clifford Johnson, Nemanja Kaloper and Rob Myers \jkkm, and the other
with Sergio Ferrara and Ruben Minasian \fkm. I would like to thank
them all for collaboration and for helpful discussions.

\vfil\eject
\listrefs
\bye